\documentstyle[aps,prb,multicol]{revtex}
\begin{document}
\draft
\preprint{Proc. OECS-5 Conference, G\"ottingen, 1997 
[To appear in Phys. Stat. Sol. (b)]}
\title{Excitonic effects in quantum wires}
\author{Guido Goldoni, Fausto Rossi, and Elisa Molinari}
\address{
Istituto Nazionale Fisica della Materia (INFM) and \\
Dipartimento di Fisica, Universit\`a di Modena \\
via Campi 213/A, I-41100 Modena, Italy
}
\date{\today}
\maketitle
\begin{abstract}
We review the effects of Coulomb correlation on the linear and
non-linear optical properties of semiconductor quantum wires, with
emphasis on recent results for the bound excitonic states.  Our
theoretical approach is based on generalized semiconductor Bloch
equations, and allows full three-dimensional multisubband description
of electron-hole correlation for arbitrary confinement profiles.  In
particular, we consider V- and T-shaped structures for which
significant experimental advances were obtained recently. Above band
gap, a very general result obtained by this approach is that
electron-hole Coulomb correlation removes the inverse-square-root
single-particle singularity in the optical spectra at band edge, in
agreement with previous reports from purely one-dimensional models.
Strong correlation effects on transitions in the continuum are found
to persist also at high densities of photoexcited carriers.

Below bandgap, we find that the same potential- (Coulomb) to
kinetic-energy ratio holds for quite different wire cross sections and
compositions. As a consequence, we identify a shape- and
barrier-independent parameter that governs a universal scaling law for
exciton binding energy with size. Previous indications that the shape
of the wire cross-section may have important effects on exciton
binding are discussed in the light of the present results.
\end{abstract}
\begin{multicols}{2}
\narrowtext

\section{Introduction}

Much of the interest in the optical properties of quantum wires was
initially motivated in terms of their single-particle properties: The
well known van Hove divergence in the one-dimensional (1D)
density-of-states (DOS) was expected to induce sharp peaks in the
optical spectra of quantum wires, thereby allowing to achieve
structures with improved optical efficiency as compared to their
two-dimensional (2D) and three-dimensional (3D) counterparts
\cite{Review_Wires}.

>From the theoretical point of view, the importance of Coulomb
correlation in determining the optical response of quantum wires was
pointed out by Ogawa and Takagahara \cite{Ogawa}. By solving a purely
1D Schr\"odinger equation in terms of a modified Coulomb potential,
they found that the $1/\sqrt{E}$ singularity at the band edge was
smoothed when electron-hole interaction was taken into account. For a
few years, however, the impact of such result was very limited. The
reason is probably twofold. On the one hand, an idealized 1D model was
thought to be far from the physics of the (wide) quantum wires that
could be actually fabricated. On the other hand, it was expected that
at high carrier densities ---the regime of interest for most
applications--- the effects of electron-hole interactions above band gap
would be washed out, in analogy with the behaviour of bound excitonic
states below band gap. From an experimental point of view, ``sharp" 1D
features at the band edge are generally not observed in the optical
spectra of quantum wires; however, owing to disorder-induced
inhomogeneous broadening, it is hard to single out the role played by
electron-hole correlation in their suppression.

An additional effort was therefore needed on the theoretical side, in
order to assess the effects of Coulomb interaction in the quantum
wires that are made available by the present technology. These include
structures obtained by epitaxial growth on non-planar substrates
(V-shaped wires) \cite{Review_Wires,Kapon,Rinaldi,Grundmann} or by
cleaved-edge quantum well overgrowth (T-shaped wires)
\cite{Wegscheider,Sakaki,Gislason}: here, the lateral extension of the
ground single-particle states is still significant (see e.g. Fig. 1), the
excited states gradually approach a 2D-like behaviour, and the subband
separation is relatively small, so that the coupling between different
subbands may be important. 

For this reason, we have recently developed \cite{PRL1,PRL2} a fully
3D approach that allows the analysis of Coulomb correlation in
realistic quantum wires, and is based on multisubband
semiconductor Bloch equations (SBE) \cite{Books}. A first set of
studies \cite{PRL1} focused precisely on the effect of electron-hole
coupling on the optical spectra above the onset of the continuum. The
main results can be summarized as follows: (i) The correlated
absorption spectra of realistic wires do show a strong quenching of
the 1D single-particle singularity, arising from the fact that the
oscillator strength vanishes when the excess energy decreases to
approach the band edge; this effect does not depend on details of the
wire cross section. The Sommerfeld factor, which is greater than unity
in the bulk and in quantum wells (the so-called Coulomb enhancement),
is instead smaller than unity in quantum wires (Coulomb suppression),
thus reducing the influence of dimensionality on the optical spectra.
In this respect, the findings of Ref.
\onlinecite{Ogawa} are therefore confirmed and their validity is extended
to realistic quantum wires with finite cross-section and arbitrary number
of subbands. (ii) The Coulomb-induced suppression of the 1D singularity is
found to hold not only in the linear regime but also at higher carrier
densities, and to persist in the gain regime. 

The above results call for a discussion of their possible implications
for perspective devices, since the initial motivations ---based on
single particle models--- are now recognized to be far too simplified.
At the same time, the theoretical approach that was developed in Ref.
\onlinecite{PRL1} can now be extended and used for an accurate study
of correlation effects below band-gap, i.e. for predicting the
properties of bound excitonic states in realistic wires \cite{PRL2}.
Beside their fundamental interest, these properties are of relevance
for possible applications in devices (e.g. nonlinear optical switches)
that would benefit of enhanced exciton binding energies.

The rest of the paper is organized as follows. In Section II we
briefly summarize the state of the art in the theoretical study of
bound excitons in quantum wires, and recall the key features of our
approach. In Section III we summarize and discuss our main results for
T-shaped and V-shaped quantum wires and compare critically with
experimental data.

\section{Bound excitons in quantum wires: theoretical background}

In the pure 1D limit, we know from the solution of the 1D hydrogenic
problem \cite{elliot-loudon} that the exciton binding energy diverges.
In real systems, where the cross section of the wire is finite,
unphysical 1D divergencies disappear because the Coulomb interaction
has a 3D nature. The resulting exciton binding energy is finite, and
its value depends on the confining potential which determines the
spatial extension of the interacting electron and hole wavefunctions.

Most of the early models for excitons in quantum wires did not rely on
a 3D description of the system: in such cases, divergencies are
avoided by assuming effective 1D potentials (usually Coulomb-like
potentials regularized at the origin) to simulate the effect of the
lateral extension of the wire. While some of these calculations have
led to the understanding of important features of correlated spectra
\cite{Ogawa,Ando} ---later confirmed by 3D studies \cite{PRL1}---,
they are intrinsically unable to account for the dependence of binding
energies on the specific confining potential. Three-dimensional
studies of Coulomb interactions in wires were also developed, mostly
based on variational approaches for the wavefunctions
\cite{Degani,Spector,Chang-Chang}.

With the advent of optical studies on V-shaped and T-shaped wires
\cite{Review_Wires,Kapon,Rinaldi,Grundmann,Wegscheider,Sakaki,Gislason}, 
several groups have
realized the importance of the specific confining potential in determining
single-particle properties of electrons and holes in the wires. Results
based on different approximations are reported e.g. in Refs.
\onlinecite{Rinaldi,Grundmann,Citrin,Goldoni} and
\onlinecite{Goldoni,Gershoni,Rossler,Langbein,Brinkmann} for V- and T-like
geometries, respectively. 

It was still unclear, however, whether the shape of the wire cross
section could influence significantly the exciton binding energies,
$E_b$. More generally, the dependence of $E_b$ on the parameters
defining the confining potential, and its scaling with size were still
highly controversial. Theoretical predictions obtained through
variational calculations for model wire geometries indicate that the
scaling of $E_b$ should be simply governed by the extension of the
single-particle wavefunctions in the plane perpendicular to the free
wire direction, and much less sensitive to the shape of the wire
\cite{Degani,Lefebvre,Reinecke}.

In view of the current debate on experimental values of $E_b$ and
their interpretation, it is useful to reexamine the influence of the
wire cross section through the SBE approach, that involves no
assumption on the form of the variational wavefuntions. The
implementation of the generalized SBE method for quantum wires
is described in full detail in
Ref.~\onlinecite{PRL1}. Here, we just stress once more that this
approach fully describes the 3D nature of the electron-hole Coulomb
interaction. Moreover, our multisubband approach includes on the same
footing both confined and continuum states; this is especially
important in order to properly describe T-wires, where the 1D ground
state is relatively close to the 2D continua of the parent QWs.


Note also that, in addition to {\it quantum-confinement effects}, the
method based on SBEs can be generalized to include {\it
dielectric-confinement effects}, i.e., the contribution due to the
dielectric-constant mismatch at the various material interfaces
\cite{PRL1}].
Contrary to some higher-symmetry systems, where these effects can be
incorporated through simple image-charge methods, for V- and T-shape
geometries their calculation require a full 3D analysis. Within our
scheme, such 3D solution is performed in the same plane-wave
representation used for the solution of the single-particle
Schr\"odinger equation. The renormalized electron-hole Coulomb
potential ---obtained as the solution of the Poisson equation with a
spatially varying dielectric constant--- is then used to compute the
optical response through the polarization equation [eq. (17) in 
Ref.~\onlinecite{PRL1}]. 
The exciton binding energy obtained in this way 
reflects both quantum- and dielectric-confinement effects. 

A more detailed account of the numerical approach for solving 
coupled SBE-Poisson equations will be given in a forthcoming publication.
Their solution is obtained by direct numerical evaluation of the polarization
eigenvalues and eigenvectors, which fully determine the absorption spectrum.
The main ingredients are the single-particle energies and wavefunctions, 
which in turn are
numerically computed (using a plane-wave expansion) starting from the
real shape of the 2D confinement potential deduced from TEM, as in
Ref. \cite{Rinaldi} (for an example see Fig. 1).

In the next Section, the results for realistic wires will be discussed
in terms of the exciton binding energy $E_b$ (calculated as the
difference between the ground-state energy of the correlated and
uncorrelated Hamiltonians), and the ground-state average of the
Coulomb and the kinetic energy, $\langle V\rangle$ and $\langle
K\rangle$. The latter is defined as the difference $\langle
K\rangle=E_b-\langle V\rangle$. Therefore, the ratio $\alpha=-\langle
V\rangle/\langle K\rangle$ is the equivalent of a virial coefficient:
in a purely 3D or 2D system, $\alpha=2$ as a direct consequence of the
virial theorem. In the presence of an additional confining potential, the
virial theorem no longer holds and $\alpha$ may deviate from 2. It is
particularly interesting to verify whether values of $\alpha>2$ can be
achieved in realistic quantum wires, as this would imply enhanced
binding energies with respect to the extreme 2D limit.

\section{Results for V- and T-shaped quantum wires}

We have considered the typical V- and T- confinement profiles
illustrated in the insets of Fig. 2, and scaled them by varying the
corresponding geometrical parameters ($L_V$ and $L_T$) in a wide range
of values. For each value of these parameters, we have considered two
sets of conduction and valence band offsets, simulating both low-$x$
Al$_x$Ga$_{1-x}$As and pure AlAs barrier compositions (samples labeled
``1" and ``2", respectively).

The calculated binding energies are illustrated by thick solid lines
in Fig. 2. As expected, they increase with decreasing $L_V$ or $L_T$;
for samples ``2", corresponding to AlAs barriers, the excitonic
binding is larger than in the case of low-barrier samples ``1". Note
the different scale of both panels: a same value of $E_b$ can be
obtained in a V- or T-shaped wire; to this purpose, the geometrical
parameters must be such that $L_T<L_V$. For an example of the
corresponding single-particle wavefunctions, see the plots in Fig. 1
which refer to the lowest circled values of $E_b$ in each panel.

The thinner lines in Fig. 2 represent the average potential energy,
$\langle V\rangle$, for the same sets of samples. The contribution of
$\langle V\rangle$ appears to dominate over $\langle K\rangle$, and to
determine the $E_b$ dependence on the geometrical parameters. It is
therefore useful to define the quantity:
\begin{equation}
a = \left\langle {1\over {\bf r}}\right\rangle^{-1},
\end{equation}
where the average is performed over the ground state. 
For a 3D bulk semiconductor, $a$ coincides with the usual exciton Bohr 
radius $a_\circ$, and can thus be considered as an effective Bohr radius
in the wire. The insets in Fig.~2 show $a$ as a function
of the relevant geometrical parameter, $L_V$ or $L_T$. A same value of $a$
corresponds to different values of $L_V$ and $L_T$, with $L_V$ always
larger than $L_T$. Note that samples with similar binding energies
correspond to similar values of $a$ (see again the circled points in Fig.
2). 

Finally, by plotting the binding energy $E_b$ of all samples vs the
corresponding exciton radius $a$ (Fig.~3), we obtain a universal
(shape and barrier independent) curve, $E_b \sim {1\over a}$. A
universal scaling of the mean potential and kinetic energy is apparent
in the $\langle V\rangle$ vs $\langle K\rangle$ plot reported in the
inset of Fig.~3; to a very good approximation, all sets of points for
V- and T-wires fall on a straight line with slope $\alpha$ very close
to $4$. For comparison, we have performed analogous calculations for a
set of QWs: we find that $E_b$ scales with $a$ similarly to wire
structures, although with a different prefactor, corresponding
approximately to $\alpha=2$ (inset of Fig.~3).

The above findings confirm that quantum wire confinement is indeed
advantageous for the purpose of obtaining enhanced exciton binding,
and provide a general and quantitative prescription for tailoring
$E_b$ by tuning the effective exciton Bohr radius $a$ through the
geometrical size parameters. At the same time, they clearly indicate
that only minor effects can be introduced by tailoring the shape of
the wire cross-section, unless smaller values of the effective Bohr
radius $a$ are obtained. This last conclusion is in apparent
contradiction with some findings that have been reported for
T-wires~\cite{Sakaki}. A reinterpretation of these data~\cite{here}
seems now to reconcile them with the present theoretical picture, as
well as with the previous experimental results \cite{Rinaldi,Weman}.

Before concluding, we briefly comment on the role of dielectric
confinement in these systems. In agreement with previous
investigations in quantum wells \cite{Andreani}, our analysis confirms
that for GaAs/AlGaAs-based structures dielectric confinement has a
minor influence as compared to quantum confinement. A slightly larger
effect, still not more than a few meV, is found for the case of the
GaAs/AlAs T-shaped structures; for example, for a symmetric GaAs/AlAs
T-wire with $L_T=5.4$ nm (sample S2 in Ref.~\onlinecite{Sakaki}), we
found a dielectric-confinement induced enhancement of $~\sim 3$ meV.
This increase is not attributed to the particular shape of the wire, 
but to the larger dielectric mismatch of the GaAs/AlAs interface and to the
smaller geometrical size of the wire.

In summary, we have presented a theoretical analysis of the optical
properties of realistic quantum wires, based on a numerical solution
of the semiconductor Bloch equations describing the multisubband 1D
system. We have applied such approach to typical T- and V-shaped
structures, whose parameters reflect the current state-of-the-art in
quantum-wire fabrication. The calculations show that exciton
binding energies in wires may indeed reach values much larger than in
quantum wells; the value of the coefficient $\alpha \sim 4$, that is found
to hold in a wide range of geometrical wire parameters, allows binding
energies larger than the ideal 2D limit. The scaling of the exciton
binding energy is governed by a universal parameter that limits the
possible differences due to variations in the shape of the wire
cross-section. Our results for realistic V- and T-wires in the strong
confinement regime are consistent with the available experimental data
and offer a guideline for tailoring binding energies in these
structures.

Acknowledgements. We are grateful to H. Sakaki for useful discussions, 
and to T.L. Reinecke for communicating results prior to publication. 

%

%
%
\begin{figure}
\caption{Single particle electron envelope functions for a V-wire with
$L_V=8.7$ nm and conduction band offset of 150 meV (left panel), 
and a T-wire with $L_T=5.4$ nm conduction band offset of 243 meV
(right panel). The exciton binding energies of these samples are the
lower circled dots in Fig.~2.}
\end{figure}

\begin{figure}
\caption{Exciton binding energy $E_b$ and mean potential energy
$\langle V\rangle$ for a set of V-wires (left) and T-wires (right).
Full dots indicate samples with pure AlAs barriers, while empty dots
indicate low-Al content samples; complete sample parameters are
summarized in Table 1 of Ref.~\protect\onlinecite{PRL2}. 
In each panel, the left insets shows the
wire geometry, with indication of the relevant geometrical parameter;
the right insets shows the calculated effective exciton Bohr radius,
$a$, vs the relevant geometrical
parameter. The circled points refer to sample parameters
corresponding to Ref.~\protect\cite{Rinaldi} (V-wires) and
Ref.~\protect\cite{Sakaki} (T-wires).}
\end{figure}
\begin{figure}
\caption{Exciton binding energy, $E_b$, vs effective exciton Bohr radius,
$a$, for V- and T-wires, and for a set of QWs.
Dashed curves are a fitting to $1/a$.
The inset shows the average potential vs kinetic energy,
falling on a straight line with slope $\alpha\simeq 4$
for all wire samples. Results for QW structures
are also shown for comparison; in this case $\alpha\simeq 2$. Solid lines
are a linear fit to the calculated points.}
\end{figure}

\end{multicols}{2}
\end{document}